\documentclass{article}
\usepackage{amssymb}
\usepackage{amsmath}

\newtheorem{theorem}{Theorem}

\newtheorem{definition}[theorem]{Definition}

\input{tcilatex}
\begin{document}

\title{Vacuum Non-Expanding Horizons and Shear-Free Null Geodesic Congruences%
}
\author{T.M. Adamo$^{1}$ \& E.T. Newman$^{2}$ \and $^{1}$Dept. of
Mathematics, University of Pittsburgh \and $^{2}$Dept. of Physics \&
Astronomy, University of Pittsburgh}
\maketitle

\begin{abstract}
We investigate the geometry of a particular class of null surfaces in
space-time called vacuum Non-Expanding Horizons (NEHs). \ Using the
spin-coefficient equation, we provide a complete description of the horizon
geometry, as well as fixing a canonical choice of null tetrad and
coordinates on a NEH. \ By looking for particular classes of null geodesic
congruences which live exterior to NEHs but have the special property that
their shear vanishes at the intersection with the horizon, a good cut
formalism for NEHs is developed which closely mirrors asymptotic theory. \
In particular, we show that such null geodesic congruences are generated by
arbitrary choice of a complex world-line in a complex four dimensional
space, each such choice induces a CR structure on the horizon, and a
particular world-line (and hence CR structure) may be chosen by transforming
to a privileged tetrad frame.
\end{abstract}

\section{Introduction}

This work is devoted to two related topics with the discussion of the second
depending on the results of the first. The associated issues are however
distinct from each other.\ The first topic is the analysis of the geometry
of certain special null 3-surfaces embedded in a four-dimensional Lorentzian
manifold. These surfaces, $\mathfrak{H}$, referred to as local non-expanding
horizons (NEHs), are defined by having the topology of $S^{2}\times \mathbb{R%
},$ with the additional property that the null generators (the null
geodesics of the surface) have both vanishing divergence and vanishing
shear. The study of local NEH geometry has a history dating back (to the
best of our knowledge) to the early vacuum work of Pajerski et al. in 1969 
\cite{Pajerski,Trapped}, where a special case of a horizon was considered.
This was followed recently by the more general and sophisticated approaches
of Ashtekar, Lewandowski and their colleagues \cite%
{Ashtekar1,Mechanics,Geometry,Lew1,Lew2}.\ In a long series of papers, using
different gauge conditions than those of Pajerski, they developed a general
theory for the intrinsic geometry of horizons for both vacuum and non-vacuum.

In the present work we return to the earlier approach of Pajerski where all
the vacuum equations of the Newman-Penrose or Spin-Coefficient (SC)
formalism are used to give a straightforward, simple derivation with
relatively transparent results of the vacuum NEH geometry. Our results also
give a complete description of the geometry for both rotating and
non-rotating non-expanding horizons in the spin-coefficient formalism.

In addition to our desire to take a second look at horizon geometry, another
major reason for our return to the topic of horizons was to investigate
certain further geometric structures that live on them which, up to now,
have been ignored. The same structures arise naturally and determine a rich
structure on the future null infinity ($\mathfrak{I}^{+}$) of asymptotically
flat space-times. Exploiting the analogous properties between $\mathfrak{I}%
^{+}$ and a NEH $\mathfrak{H}$ (e.g., both $\mathfrak{I}^{+}$ and $\mathfrak{%
H}$ are $S^{2}\times \mathbb{R}$ null surfaces with the null generators
having vanishing divergence and shear), we can generate these same
structures on $\mathfrak{H}$. \ 

More precisely, we study Null Geodesic Congruences (NGCs) "living" exterior
to $\mathfrak{H}$ but that intersect $\mathfrak{H}$ with the very special
property that their shear vanishes at $\mathfrak{H}$. We will refer to such
congruences as $\mathfrak{H}$-shear-free\ (this is the analogue of
asymptotically shear-free NGCs). In order to do this, the geometry of $%
\mathfrak{H}$ must be locally complexified so that the $\mathfrak{H}$
coordinates become complex variables slightly extended away from their real
values.

The basic result is that the $\mathfrak{H}$-shear-free congruences are
determined by solutions to the $\mathfrak{H}$-Good Cut equation, whose
solution space is a four complex dimensional manifold, a space similar to
the one that arises while studying asymptotically shear-free NGCs on $%
\mathfrak{I}^{+}$ \cite{GCE}. In the same manner as in the asymptotic case,\
any arbitrary analytic world-line in this complex space generates an $%
\mathfrak{H}$-shear-free NGC just as in the recently developed physical
identification theory on $\mathfrak{I}^{+}$ (c.f., \cite%
{UCF,PhysicalContent,Review}). \ Among these arbitrary world-lines there is
a means of singling out a unique one from which there is hope of developing
a physical identification theory on NEHs in further analogy with the more
recent work on $\mathfrak{I}^{+}$.

In Section 2, we characterize the geometry of vacuum NEHs by first defining
a coordinate and null tetrad system, then integrating the spin-coefficient
equations on the horizon. This will include the repeated use of gauge
freedoms which involve the choice of coordinates and tetrad and allow us to
fix these systems completely.\ In Section 3 we find a good cut equation that
characterizes null geodesic congruences whose shear vanishes at the horizon,
and show that such NGCs are generated by arbitrary world-lines in a complex
four-dimensional space. \ It is also observed that each choice of such a
world-line induces a CR structure on the horizon -\textbf{\ }the unique
world-line induces a unique CR structure.\ Section 4 concludes and discusses
the results, while an appendix provides a more detailed exposition of the
horizon CR structures.

\section{Non-Expanding Vacuum Horizon Geometry}

In the remainder of this paper, we will be working on a generic
non-expanding horizon, as defined by Ashtekar et al. \cite{Geometry}:

\begin{definition}
A non-expanding vacuum horizon $\mathfrak{H}$ is a null 3-submanifold in a
space-time $\mathcal{M}$ which satisfies the following properties:

(1). \ $\mathfrak{H}$ is topologically $\mathbb{R}\times S^{2}$, and there
is a projection $\Pi :\mathfrak{H}\rightarrow S^{2}$ where the fibers of
this projection are null curves in $\mathfrak{H}$;

(2). \ The complex divergence and shear of any null tangent vector $l$ to $%
\mathfrak{H}$ vanish; and

(3). \ The vacuum Einstein field equations hold on $\mathfrak{H}$.
\end{definition}

A substantial body of research had been dedicated towards understanding the
intrinsic geometry induced on NEHs by the global geometry of $\mathcal{M}$;
the interested reader may reference \cite{Mechanics,Geometry,Lew1,Lew2} for
a discussion of these findings. \ 

Working with the Spin-Coefficient (SC) formalism in the following section we
re-investigate this issue.

\subsection{Coordinates and Null Tetrads}

Consider a region in the space-time $\mathcal{R}\subset \mathcal{M}$ which
is foliated by null surfaces, each of the leaves of this foliation having
topology $\mathbb{R}\times S^{2}$. \ On $\mathcal{R}$, we choose a
coordinate $s$ to label each of these null surfaces (i.e., $s=const.$
determines a null surface with topology $\mathbb{R}\times S^{2}$). \ Each
constant $s$ slice can then be charted with coordinates $(u,\zeta ,\bar{\zeta%
})$, where $u$ covers the $\mathbb{R}$ portion of the topology and provides
a foliation of each null surface, and $(\zeta ,\bar{\zeta})$ charts the
2-sphere. \ We take the complex coordinate $\zeta =e^{i\phi }\cot (\theta
/2) $ ($\zeta =x+iy$) to be the usual complex stereographic angle coordinate
on $S^{2}$. \ Hence, we cover all of $\mathcal{R}$ with the coordinate
system $(u,s,\zeta ,\bar{\zeta})$.

Initially, we have the freedom $s\rightarrow s^{\ast }=G(s)$ and $%
u\rightarrow u^{\ast }=F(u,s,\zeta ,\bar{\zeta})$, but we will impose
coordinate and null tetrad conditions through the course of this paper which
fix the choice of these coordinates entirely.

In addition we construct a null tetrad system $\{l,n,m,\bar{m}\}$ on $%
\mathcal{R}$ which obeys the usual inner product relations:%
\begin{equation}
l^{a}n_{a}=-m^{a}\bar{m}_{a}=1,  \label{IP}
\end{equation}%
with all other contractions between the vectors vanishing. \ We set the null
vector $l$ to be the future directed tangent vector to the constant $s$ null
surfaces:%
\begin{equation}
l=l^{a}\frac{\partial }{\partial x^{a}}=\frac{\partial }{\partial u},\ \
l_{a}dx^{a}=ds.  \label{l1}
\end{equation}%
By also demanding that $u$ be an affine parameter for null geodesics on
these null surfaces, we reduce the remaining coordinate freedom in $u$ to%
\begin{equation}
u\rightarrow u^{\ast }=Z(s,\zeta ,\bar{\zeta})u+A(s,\zeta ,\bar{\zeta}),
\label{ut1}
\end{equation}%
and also fix $l$ as a geodesic tangent vector on the leaves of the $s$%
-foliation. \ The most general form for the remaining tetrad vectors under
these conditions is:%
\begin{eqnarray}
l &=&l^{a}\frac{\partial }{\partial x^{a}}=\frac{\partial }{\partial u},\ \
l_{a}dx^{a}=ds,  \label{tet1} \\
n &=&n^{a}\frac{\partial }{\partial x^{a}}=U\frac{\partial }{\partial u}+%
\frac{\partial }{\partial s}+X^{A}\frac{\partial }{\partial x^{A}},
\label{tet*} \\
m &=&m^{a}\frac{\partial }{\partial x^{a}}=\omega \frac{\partial }{\partial u%
}+\xi ^{A}\frac{\partial }{\partial x^{A}},  \label{tet**}
\end{eqnarray}%
where $A=\{2,3\}$, $(x^{2},x^{3})=(\zeta $, $\bar{\zeta})$, and $U,$ $%
X^{A}\in \mathbb{R}$, $\omega ,\xi ^{A}\in \mathbb{C}$ are functions to be
determined.

We choose one particular surface, labeled by $s=0$, in $\mathcal{R}$ as our
non-expanding horizon as in Definition (1). \ Our `$s$' freedom is then:%
\begin{eqnarray}
s &\rightarrow &s^{\ast }=G(s),\ \ G(0)=0.  \label{strans} \\
G^{\prime }(0) &=&\text{constant}=K\text{.}
\end{eqnarray}%
On the horizon this means that both $l$ and $n$ are trivially rescaled by a
constant. \ Our coordinate system singles out a NEH, $\mathfrak{H}$, in the
region $\mathcal{R}$ as:%
\begin{equation}
\mathfrak{H}=\{x^{a}\in \mathcal{R}:s=0\}.  \label{Horiz1}
\end{equation}

Restricting ourselves to $\mathfrak{H}$, our remaining freedom in the choice
of coordinates is:%
\begin{eqnarray}
u^{\ast } &=&Z(\zeta ,\overline{\zeta })u+A(\zeta ,\bar{\zeta}),
\label{trans} \\
s^{\ast } &=&\text{\ }sK,  \label{l hat} \\
\zeta ^{\ast } &=&\frac{a\zeta +b}{c\zeta +d},\text{ \ }ad-bc=1,  \label{su2}
\end{eqnarray}

\noindent the four-parameters $\{a,b,c,d\}$ lie in SL$(2,\mathbb{C})$; $%
\zeta \rightarrow \zeta ^{\ast }$ being the fractional linear transformation
that maps $S^{2}$ to itself. \ The remaining freedom in the $s$ coordinate
will simply re-scale the null tetrad by $K$ and can be bundled into the $Z$
for the $u$-transformations, which will be fixed later. \ Under the action
of the transformation Eq.(\ref{trans}), it should be noted that%
\begin{equation*}
l\rightarrow l^{\ast }=Zl,
\end{equation*}%
so if we demand that 
\begin{equation*}
l=\frac{\partial }{\partial u}\text{, \ }l^{\ast }=\frac{\partial }{\partial
u^{\ast }},
\end{equation*}%
it follows that any transformation of the $u$ coordinate must be accompanied
by a re-scaling of the null tetrad as: 
\begin{equation}
l^{\ast }=Z^{-1}l,  \label{rescale l}
\end{equation}%
along with 
\begin{equation}
n^{\ast }=Zn,  \label{rescale2}
\end{equation}%
to preserve the inner product $l^{a}n_{a}=1$.

The full four-dimensional (contravariant) metric evaluated with this tetrad
is given by%
\begin{equation}
g^{ab}=l^{a}n^{b}+n^{a}l^{b}-m^{a}\overline{m}^{b}-\overline{m}^{a}m^{b}.
\label{4metric}
\end{equation}

\noindent Without going into details since they are already in the
literature \cite{NewmanTod,NPF} we outline another coordinate condition that
greatly simplifies the analysis. The metric given by Eq.(\ref{4metric})
induces a 2-metric on the 2-surfaces $u=$ \textit{constant} on $\mathfrak{H}$
($s=0$) that is determined by the coefficients $\xi ^{A}$ appearing in the
tetrad. Since every 2-metric is conformally flat, coordinates on the
2-surface can be introduced so that the metric has the form 
\begin{equation}
ds^{2}=P^{-2}d\zeta d\overline{\zeta }.  \label{2metric}
\end{equation}%
This in turn, using the freedom of a spin-transformation $m\rightarrow
m^{\ast }=e^{i\theta }m,$ allows the $\xi ^{A}$ to be chosen, with $P$ real,
as:%
\begin{align}
\xi ^{\zeta }& =-P,\ \xi ^{\overline{\zeta }}=0,\ \ \ \   \label{P*} \\
\ \ \ \ \ \overline{\xi }^{\zeta }& =0,\text{ \ \ }\overline{\xi }^{%
\overline{\zeta }}=-P.  \notag
\end{align}%
This form of $\xi ^{A}$ is used throughout this work.

\textbf{Note: }The form of Eq.(\ref{tet**}) simplifies to%
\begin{equation}
m=\omega \frac{\partial }{\partial u}-P\frac{\partial }{\partial \zeta }.
\label{tet***}
\end{equation}

\textbf{Remark \ }Note that although the metric, (\ref{2metric}), is
conformal to a sphere metric, it need not be a sphere metric itself. \
Hence, we will often write:%
\begin{equation}
P=VP_{0},  \label{conformal}
\end{equation}%
where $V$ is the conformal factor and $P_{0}$ induces a 2-sphere metric. \
This has important implications for the definition of the $\eth $-operator
on $\mathfrak{H}$, which will prove crucial later. \ For a spin-weight $s$
function $f_{(s)}$ defined on $\mathfrak{H}$, we have:%
\begin{eqnarray}
\eth f_{(s)} &=&P^{1-s}\frac{\partial }{\partial \zeta }(P^{s}f_{(s)}),
\label{horedth} \\
\bar{\eth }f_{(s)} &=&P^{1+s}\frac{\partial }{\partial \bar{\zeta}}%
(P^{-s}f_{(s)}),  \notag
\end{eqnarray}%
and we will at some points need to compare this to the $\eth _{0}$-operator
defined on the 2-sphere as%
\begin{eqnarray}
\eth _{0}f_{(s)} &=&P_{0}^{1-s}\frac{\partial }{\partial \zeta }%
(P_{0}^{s}f_{(s)}),  \label{edth} \\
\bar{\eth }_{0}f_{(s)} &=&P_{0}^{1+s}\frac{\partial }{\partial \bar{\zeta}}%
(P_{0}^{-s}f_{(s)}).  \notag
\end{eqnarray}%
In particular, to work with tensorial spin-$s$ spherical harmonics in
equations that contain $\eth $, we must make all expressions in terms of $%
\eth _{0}$. \ This will be particularly important when developing a good cut
formalism for non-expanding horizons. $\blacksquare $

Returning to the discussion of the gauge freedom, we have the choice of null
rotations about the $l$ vector:

\begin{eqnarray}
l &\rightarrow &l^{\ast }=l,  \label{nullrotl} \\
m &\rightarrow &m^{\ast }=m+Ll,  \notag \\
n &\rightarrow &n^{\ast }=n+\overline{L}m+L\bar{m}+L\overline{L}l,  \notag
\end{eqnarray}%
with $L$ an arbitrary spin-weight one function. This preserves the form of
Eq.(\ref{tet1}). \ In the following section this freedom is used
extensively. \ The analogous null rotation about $n$ is not of use since it
would destroy the tangency conditions placed on $l$ in the region $\mathcal{R%
}$. \ However the rescaling freedom (boosts), $l^{\ast }=Z^{-1}l,$ $n^{\ast
}=Zn$ is used later.

In the following sub-section further conditions are placed on both the
coordinates and tetrad so that the choice of $u$ is fixed and the tetrad is
made unique.

\subsection{The Spin-Coefficients}

In this sub-section, we write down all of the Spin-Coefficient (SC)
equations on the vacuum NEH $\mathfrak{H}$, integrating many to obtain the
full $u$-dependence for the SCs. \ We also simplify these results using the
remaining gauge freedom of (\ref{trans}). \ Since nearly all of our
calculations are performed on $\mathfrak{H}$, we will omit notation such as $%
f|_{\mathfrak{H}}=f|_{s=0}$, and instead just write $f$; if the off-horizon
variable $s$ does enter, it will be stated so explicitly.

Before writing down the SC equations, we note that many of the SCs can be
restricted or made to vanish \textit{a priori} simply by conditions placed
on the null tetrad in the previous section. \ First of all, the choice of $u$
as a geodesic parameter, along with the requirement that the vectors $\{n,m,%
\bar{m}\}$ be parallely propagated along the $l$-congruence results in \cite%
{NPF}:

\textit{Tetrad Condition I:}%
\begin{equation}
\kappa =\varepsilon =\pi =0,  \label{tc1}
\end{equation}%
which in fact holds not just on $\mathfrak{H}$ but everywhere in $\mathcal{R}
$. \ There remains the freedom of the initial choice on the $n$ and $m$
vectors before their parallel propagation. \textbf{\ }This permits the
rotation parameter $L$ in Eq.(\ref{nullrotl}) to remain a free function of ($%
\zeta ,\overline{\zeta }$). This will prove to be useful later.

Next, by (\ref{l1}), we have that $l$ is a gradient vector field; this
follows from $l_{a}dx^{a}=ds$ and results in

\textit{Tetrad Condition II:}%
\begin{equation}
\tau =\bar{\alpha}+\beta ,  \label{tc2}
\end{equation}%
which also holds everywhere in $\mathcal{R}$. \ 

Finally, by the definition of $\mathfrak{H}$ as a non-expanding horizon (see
Definition 1 above), it follows that the (complex) divergence and shear of
any null tangent to the horizon must vanish. \ Hence, we have

\textit{Tetrad Condition III:}%
\begin{equation}
\rho =\sigma =0,  \label{tc3}
\end{equation}%
holding on $\mathfrak{H}$.

In addition to these three tetrad conditions, it also follows from the
Goldberg-Sachs Theorem (or the SC equations themselves) that the Weyl tensor
components vanish on the horizon as well%
\begin{equation}
\psi _{0}=\psi _{1}=0.  \label{GST}
\end{equation}%
\ We thus have, from the start, that:%
\begin{eqnarray}
\kappa &=&\varepsilon =\pi =\tau -\bar{\alpha}-\beta =\rho =\sigma =0,
\label{SCP} \\
\psi _{0} &=&\psi _{1}=0,  \notag
\end{eqnarray}%
on $\mathfrak{H}$.

To write down the full set of SC equations, we define the differential
operators:%
\begin{eqnarray*}
D &\equiv &\frac{\partial }{\partial u}, \\
\delta &\equiv &\omega \frac{\partial }{\partial u}+\xi ^{A}\frac{\partial }{%
\partial x^{A}}, \\
\Delta &\equiv &\frac{\partial }{\partial s}+U\frac{\partial }{\partial u}%
+X^{A}\frac{\partial }{\partial x^{A}}.
\end{eqnarray*}%
The spin-coefficient equations on $\mathfrak{H}$ are separated into three
sets: (1.) those containing the operator $D$ and $\delta $ or $\overline{%
\delta },$ (2.) those containing only $\delta $ and or $\overline{\delta }$
and (3.) those that contain $\Delta $ with other derivatives.$.$ The
procedure is to first integrate the $D$ equations, which determines the $u$
behavior, and then substitute those results into the second set which yields
relationships between the integration "constants" from the first set. The
third set - included for completeness - then would yield the $s$%
-derivatives, the derivatives off $\mathfrak{H}$, of a variety of the
variables. They are not of interest to us here. \ We present each set of
equations in three blocks: those for the spin-coefficients themselves, those
for the Weyl tensor, and those for the metric coefficients. \ Note that
where convenient, we have used the $\eth $-operator.

\textit{The }$D$\textit{-equations:}%
\begin{eqnarray}
D\tau &=&0,  \label{SCD} \\
D\alpha &=&0,  \notag \\
D\beta &=&0,  \notag \\
D\gamma &=&\tau \alpha +\bar{\tau}\beta +\psi _{2},  \notag \\
D\lambda &=&0,  \notag \\
D\mu &=&\psi _{2},  \notag \\
D\nu &=&\bar{\tau}\mu +\tau \lambda +\psi _{3},  \notag
\end{eqnarray}%
\begin{eqnarray}
D\psi _{2} &=&0,  \label{WTD} \\
D\psi _{3} &=&-\bar{\eth }\psi _{2},  \notag \\
D\psi _{4} &=&-\bar{\eth }\psi _{3}-3\lambda \psi _{2}-\bar{\omega}\bar{\eth 
}\psi _{2},  \notag
\end{eqnarray}%
\begin{eqnarray}
DU &=&-(\gamma +\bar{\gamma})+\bar{\omega}\tau +\omega \bar{\tau},
\label{MCD} \\
DX^{A} &=&\tau \bar{\xi}^{A}+\bar{\tau}\xi ^{A},  \notag \\
D\omega &=&-\tau ,  \notag
\end{eqnarray}

\textit{The (}$\delta $, $\bar{\delta})$-equations:%
\begin{eqnarray}
\delta \alpha -\bar{\delta}\beta &=&\alpha \bar{\alpha}+\beta \bar{\beta}%
-2\alpha \beta -\psi _{2},  \label{SCd} \\
\delta \lambda -\bar{\delta}\mu &=&\mu (\alpha +\bar{\beta})+\lambda (\bar{%
\alpha}-3\beta )-\psi _{3},  \notag
\end{eqnarray}%
\begin{eqnarray}
\bar{\delta}\omega -\delta \bar{\omega} &=&(\mu -\bar{\mu})-\tau \bar{\omega}%
+\bar{\tau}\omega ,  \label{MCd} \\
\bar{\delta}\xi ^{A}-\delta \bar{\xi}^{A} &=&-\tau \bar{\xi}^{A}+\bar{\tau}%
\xi ^{A}.  \notag
\end{eqnarray}

\textit{The }$\Delta $\textit{-equations }(Note: unless a quantity vanishes
on all of $\mathcal{R}$, it is not necessarily true that its $\Delta $%
-derivative on $\mathfrak{H}$ will vanish!)\textit{:}%
\begin{eqnarray}
\Delta \lambda -\bar{\delta}\nu &=&-(\mu +\bar{\mu})\lambda -(3\gamma -\bar{%
\gamma})\lambda +2\alpha \nu -\psi _{4},  \label{SCT} \\
\delta \nu -\Delta \mu &=&(\mu ^{2}+\lambda \bar{\lambda})+(\gamma +\bar{%
\gamma})\mu -2\beta \nu ,  \notag \\
\delta \gamma -\Delta \beta &=&\mu \tau -\beta (\gamma -\bar{\gamma}-\mu
)+\alpha \bar{\lambda},  \notag \\
\delta \tau -\Delta \sigma &=&(\tau +\beta -\bar{\alpha})\tau ,  \notag \\
\Delta \rho -\bar{\delta}\tau &=&(\bar{\beta}-\alpha -\bar{\tau})\tau -\psi
_{2},  \notag \\
\Delta \alpha -\bar{\delta}\gamma &=&-(\tau +\beta )\lambda +(\bar{\gamma}-%
\bar{\mu})\alpha +(\bar{\beta}-\bar{\tau})\gamma -\psi _{3},  \notag
\end{eqnarray}%
\begin{eqnarray}
\Delta \psi _{0} &=&0,  \label{WTT} \\
\Delta \psi _{1} &=&-\eth \psi _{2},  \notag \\
\Delta \psi _{2} &=&\delta \psi _{3}-3\mu \psi _{2},  \notag \\
\Delta \psi _{3} &=&\delta \psi _{4}+3\nu \psi _{2}-2(\gamma +2\mu )\psi
_{3},  \notag
\end{eqnarray}%
\begin{eqnarray}
\delta U-\Delta \omega &=&-\nu +\bar{\lambda}\bar{\omega}+(\mu -\gamma +\bar{%
\gamma})\omega ,  \label{MCT} \\
\delta X^{A}-\Delta \xi ^{A} &=&\bar{\lambda}\bar{\xi}^{A}+(\mu -\gamma +%
\bar{\gamma})\xi ^{A}.  \notag
\end{eqnarray}

The sets of $D$-equations are easily integrated to give the full $u$%
-dependence of all of the variables on $\mathfrak{H}$:%
\begin{eqnarray}
\tau &=&\tau _{0},  \label{SC1} \\
\alpha &=&\alpha _{0},  \notag \\
\beta &=&\beta _{0},  \notag \\
\gamma &=&\gamma _{0}+u(\tau _{0}\alpha _{0}+\bar{\tau}_{0}\beta _{0}+\psi
_{2,0}),  \notag \\
\lambda &=&\lambda _{0},  \notag \\
\mu &=&\mu _{0}+u\psi _{2,0},  \notag \\
\nu &=&\nu _{0}+u(\bar{\tau}_{0}\mu _{0}+\tau _{0}\lambda _{0}+\psi _{3,0})+%
\frac{u^{2}}{2}(\bar{\tau}_{0}\psi _{2,0}-\bar{\eth }\psi _{2,0}),  \notag
\end{eqnarray}%
\begin{eqnarray}
\psi _{2} &=&\psi _{2,0},  \label{WT1} \\
\psi _{3} &=&\psi _{3,0}-u\bar{\eth }\psi _{2,0},  \notag \\
\psi _{4} &=&\psi _{4,0}-u\left( \bar{\eth }\psi _{3,0}+3\lambda _{0}\psi
_{2,0}+\bar{\omega}_{0}\bar{\eth }\psi _{2,0}\right) +\frac{u^{2}}{2}\left( 
\bar{\eth }^{2}\psi _{2,0}+\bar{\tau}_{0}\bar{\eth }\psi _{2,0}\right) , 
\notag
\end{eqnarray}%
\begin{eqnarray}
U &=&U_{0}+u(\tau _{0}\bar{\omega}_{0}+\bar{\tau}_{0}\omega _{0}-\gamma _{0}-%
\overline{\gamma }_{0})-\frac{u^{2}}{2}(4\tau _{0}\overline{\tau }_{0}+\psi
_{2,0}+\overline{\psi }_{2,0}),  \label{MC1} \\
X^{A} &=&X_{0}^{A}-u(\tau _{0}\bar{\xi}^{A}+\bar{\tau}_{0}\xi ^{A}),  \notag
\\
\omega &=&\omega _{0}-u\tau _{0},  \notag
\end{eqnarray}%
where a subscript $f_{0}$ indicates that $f$ is a function only on the
2-sphere (i.e., $f_{0}=f_{0}(\zeta ,\bar{\zeta})$). \ 

Our procedure is now to feed these relations into the second set of SC
equations, i.e., the ($\delta ,\overline{\delta }$) equations:

After a straightforward but slightly tedious calculation we obtain from Eqs.(%
\ref{SCd}) and (\ref{MCd}) the relations:%
\begin{eqnarray}
\alpha _{0} &=&\frac{\bar{\tau}_{0}}{2}-\frac{1}{2}\frac{\partial P}{%
\partial \bar{\zeta}},  \label{ab} \\
\beta _{0} &=&\frac{\tau _{0}}{2}+\frac{1}{2}\frac{\partial P}{\partial
\zeta },  \notag
\end{eqnarray}%
\begin{equation}
\psi _{2,0}=\frac{1}{2}\left( \eth \bar{\tau}_{0}-\bar{\eth }\tau
_{0}\right) -\left( P\partial _{\zeta }\partial _{\bar{\zeta}}P-\partial _{%
\bar{\zeta}}P\cdot \partial _{\zeta }P\right) ,  \label{psi2}
\end{equation}%
\begin{equation}
\eth \bar{\omega}_{0}-\bar{\eth }\omega _{0}=\mu _{0}-\bar{\mu}_{0}+\bar{%
\omega}_{0}\tau _{0}-\omega _{0}\bar{\tau}_{0}+u\left( \psi _{2,0}-\bar{\psi}%
_{2,0}-\bar{\eth }\tau _{0}+\eth \bar{\tau}_{0}\right) ,  \label{ommu}
\end{equation}%
\begin{equation}
\psi _{3,0}=\eth \lambda _{0}-\tau _{0}\lambda _{0}-\overline{\eth }\mu
_{0}+\mu _{0}\overline{\tau }_{0}.  \label{psi3}
\end{equation}

From Eq.(\ref{psi2}) and the reality of $P$ we have:%
\begin{equation}
2\func{Im}(\psi _{2,0})=\psi _{2,0}-\overline{\psi }_{2,0}=\eth \overline{%
\tau }_{0}-\overline{\eth }\tau _{0}.  \label{Real1}
\end{equation}%
This allows us to define a mass aspect on $\mathfrak{H}$, analogous to the
Bondi mass aspect on $\mathfrak{I}^{+}$,%
\begin{equation}
\Psi \equiv \psi _{2,0}+\overline{\eth }\tau _{0}=\bar{\Psi}.  \label{MA}
\end{equation}%
Additionally, recall that the full space-time metric $g^{ab}$, when pushed
down to the $u=const.$ 2-surfaces of $\mathfrak{H}$, induces a 2-metric with
line element:%
\begin{equation*}
ds^{2}=P^{-2}d\zeta d\bar{\zeta}.
\end{equation*}%
The scalar curvature of this 2-surface, which is topologically $S^{2}$ for
each fixed $u$, is given by:%
\begin{equation}
K=2(P\partial _{\zeta }\partial \overline{_{\zeta }}P-\partial \overline{%
_{\zeta }}P\partial _{\zeta }P),  \label{SC}
\end{equation}%
so that%
\begin{equation}
\psi _{2,0}=\frac{1}{2}\left[ -K+\left( \eth \bar{\tau}_{0}-\bar{\eth }\tau
_{0}\right) \right] .  \label{psi2*}
\end{equation}

\textit{Tetrad Condition IV:}

Using the remaining gauge freedom of the ($\zeta ,\overline{\zeta }$%
)-dependent null rotation about the vector $l$ from (\ref{nullrotl}), and
choosing $L$ to have the form%
\begin{equation*}
L(\zeta ,\bar{\zeta})=-\omega _{0},
\end{equation*}%
the vector $m$ transforms to:%
\begin{eqnarray}
m^{\ast } &=&(\omega _{0}-u\tau _{0})\frac{\partial }{\partial u}-P\frac{%
\partial }{\partial \zeta }-\omega _{0}\frac{\partial }{\partial u}
\label{mtrans} \\
&=&-u\tau _{0}\frac{\partial }{\partial u}-P\frac{\partial }{\partial \zeta }%
.  \notag
\end{eqnarray}%
so the new $\omega _{0}^{\ast }$ vanishes. \ Dropping the "$^{\ast }$" we
have that%
\begin{equation}
\omega _{0}=0.  \label{omega=0}
\end{equation}

Using these results, i.e., Eq.(\ref{omega=0}) and (\ref{MA}), in Eq.(\ref%
{ommu}), we finally have that%
\begin{equation}
\mu _{0}=\bar{\mu}_{0}.  \label{ommu*}
\end{equation}%
\ The null rotation freedom is now fixed.

To summarize all of our results thus far, we have:%
\begin{eqnarray}
\tau  &=&\tau _{0},  \label{scf1} \\
\alpha  &=&\alpha _{0}=\frac{1}{2}\overline{\tau }_{0}-\frac{1}{2}\partial _{%
\overline{\zeta }}P, \\
\beta  &=&\beta _{0}=\frac{1}{2}\tau _{0}+\frac{1}{2}\partial _{\zeta }P, \\
\lambda  &=&\lambda _{0}, \\
\mu  &=&\mu _{0}+u\psi _{2,0},\text{ \ \ }\mu _{0}=\overline{\mu }_{0}, \\
\nu  &=&\nu _{0}+u(\tau _{0}\lambda _{0}+\bar{\tau}_{0}\mu _{0}+\psi _{3,0})+%
\frac{u^{2}}{2}\left( \bar{\tau}_{0}\psi _{2,0}-\bar{\eth }\psi
_{2,0}\right) , \\
\gamma  &=&\gamma _{0}+u(\tau _{0}\alpha _{0}+\bar{\tau}_{0}\beta _{0}+\psi
_{2,0}) \\
&=&\gamma _{0}+u\left( \tau _{0}\overline{\tau }_{0}-\frac{\tau _{0}}{2}%
\partial _{\overline{\zeta }}P+\frac{\bar{\tau}_{0}}{2}\partial _{\zeta
}P+\psi _{2,0}\right) , \\
\omega  &=&-\tau _{0}u, \\
\xi ^{A} &=&\xi _{0}^{A}=-(P,0),\ \ \overline{\xi }^{A}\ =\overline{\xi }%
_{0}^{A}=-(0,P), \\
U &=&U_{0}-u(\gamma _{0}+\overline{\gamma }_{0})-\frac{u^{2}}{2}(4\tau _{0}%
\overline{\tau }_{0}+\psi _{2,0}+\overline{\psi }_{2,0}), \\
X^{A} &=&X_{0}^{A}+u(\tau _{0}\overline{\xi }_{0}^{A}+\overline{\tau }%
_{0}\xi _{0}^{A}), \\
\psi _{2} &=&\psi _{2,0}=\frac{1}{2}\left[ -K+\left( \eth \bar{\tau}_{0}-%
\bar{\eth }\tau _{0}\right) \right] , \\
\text{\ \ \ \ \ \ }K &=&2(P\partial _{\zeta }\partial \overline{_{\zeta }}%
P-\partial \overline{_{\zeta }}P\partial _{\zeta }P), \\
\text{ \ \ }\Psi  &=&\psi _{2,0}+\overline{\eth }\tau _{0}=\overline{\Psi }=-%
\frac{1}{2}K+\frac{1}{2}\eth \overline{\tau }_{0}+\frac{1}{2}\overline{\eth }%
\tau _{0}, \\
\psi _{3} &=&\psi _{3,0}-u\bar{\eth }\psi _{2,0}, \\
\psi _{3,0} &=&\eth \lambda _{0}-\tau _{0}\lambda _{0}-\overline{\eth }\mu
_{0}+\mu _{0}\overline{\tau }_{0}, \\
\psi _{4} &=&\psi _{4,0}-u\left( \bar{\eth }\psi _{3,0}+3\lambda _{0}\psi
_{2,0}+\bar{\eth }\psi _{2,0}\right) +\frac{u^{2}}{2}\left( \bar{\eth }%
^{2}\psi _{2,0}+\bar{\tau}_{0}\bar{\eth }\psi _{2,0}\right) .  \label{scf}
\end{eqnarray}

We conclude this section by exploiting the remaining coordinate freedom in
the choice of the origin of the $u$ coordinate to further simplify the
spin-coefficient results. \ Recall from (\ref{trans}) that the remaining
freedom in $u$ is:%
\begin{equation*}
u\rightarrow u^{\ast }=Z(\zeta ,\bar{\zeta})u+A(\zeta ,\bar{\zeta}),
\end{equation*}%
where "boosts" are given with the $Z$ and "supertranslations" with the $A$.
\ We use the boost freedom to make the spin-coefficient $\tau $ "pure
magnetic," while the supertranslation freedom will be used to eliminate $\mu
_{0}$ altogether.

Recall that under a boost $Z(\zeta ,\bar{\zeta})$, the tetrad vectors $l$
and $n$ must be rescaled as:%
\begin{eqnarray*}
l &\rightarrow &l^{\ast }=Z^{-1}l, \\
n &\rightarrow &n^{\ast }=Zn,
\end{eqnarray*}%
in order to preserve our tetrad conditions. \ If we write 
\begin{equation}
Z^{-1}\equiv F(\zeta ,\bar{\zeta}),  \label{transF}
\end{equation}%
then the spin-coefficient $\tau $ transforms under the boost $Z$ as:%
\begin{equation}
\tau \rightarrow \tau ^{\ast }=\tau +F^{-1}\eth F.  \label{tau1}
\end{equation}%
Let us assume that the spin-weight one function $\tau =\tau _{0}$ is smooth
enough that it can be written as%
\begin{equation}
\tau _{0}=\eth \text{\textsc{t}}=\eth (\text{\textsc{t}}_{R}+i\text{\textsc{t%
}}_{I}),  \label{tau2}
\end{equation}%
where \textsc{t} is a holomorphic spin-weight zero function on the 2-sphere.
\ By choosing:%
\begin{eqnarray}
Z &=&e^{\text{\textsc{t}}_{R}},  \label{tau3} \\
F &=&e^{-\text{\textsc{t}}_{R}},  \notag
\end{eqnarray}%
we see that%
\begin{eqnarray}
\tau _{0}^{\ast } &=&\tau _{0}+F^{-1}\eth F=\tau _{0}+\eth (\log F)
\label{tau4} \\
&=&\tau _{0}-\eth \text{\textsc{t}}_{R}=\eth (\text{\textsc{t}}_{R}+i\text{%
\textsc{t}}_{I}-\text{\textsc{t}}_{R}).  \notag
\end{eqnarray}%
Hence, with this choice of boost for the $u$ gauge, we see that $\tau _{0}$
is a pure "magnetic" type function:%
\begin{equation}
\tau _{0}^{\ast }=i\eth \text{\textsc{t}}_{I}.  \label{tau5}
\end{equation}%
Note that this results remain true even though $\eth $ is different from $%
\eth _{0}.$

Finally, we fix the supertranslation freedom by considering the
spin-coefficient $\mu $. \ Recall that for%
\begin{equation*}
\mu =\mu _{0}+u\psi _{2,0},
\end{equation*}%
we have already established that $\mu _{0}\in \mathbb{R}$, so it follows
that:%
\begin{eqnarray}
\func{Re}(\mu ) &\equiv &\mu _{R}=\mu _{0}+u\func{Re}(\psi _{2,0})
\label{mu1} \\
&=&\mu _{0}-\frac{u}{2}K,  \notag
\end{eqnarray}%
where $K$ is the scalar curvature of $u=const.$ cross sections given by
equation (\ref{SC}). \ Under a supertranslation with%
\begin{eqnarray}
u^{\ast } &=&u+A  \label{mu2} \\
A(\zeta ,\bar{\zeta}) &=&\frac{2}{K}\mu _{0},
\end{eqnarray}%
the $u$ origin is shifted to the "cut" where $\mu _{R}$ vanishes. This in
turn sets%
\begin{equation}
\mu _{0}^{\ast }=0.  \label{mu4}
\end{equation}

To summarize, we have now totally fixed the choice of tetrad and coordinate
gauge (except for fractional linear transformations on $\zeta $), resulting
in the conditions (dropping the "$^{\ast }$" notation) that:%
\begin{eqnarray}
\omega _{0} &=&0,  \label{GCs} \\
\tau _{0} &=&i\eth \text{\textsc{t}}_{I},  \notag \\
\mu _{0} &=&0.  \notag
\end{eqnarray}%
These in turn lead to the simplification of several of the expressions in (%
\ref{scf1})-(\ref{scf}), particularly:%
\begin{equation*}
\psi _{3,0}=\eth \lambda _{0}-\tau _{0}\lambda _{0}.
\end{equation*}

The remaining free functions of $(\zeta ,\bar{\zeta})$ are:%
\begin{eqnarray*}
\text{complex}\text{: } &&\tau _{0},\lambda _{0},\gamma _{0},\nu _{0},\psi
_{4,0}, \\
\text{real} &\text{:}&\text{ }P,U_{0},X_{0}^{A}.
\end{eqnarray*}%
A particularly special simple case is when $P=P_{0},$ i.e., when the
2-metric is that of a sphere.

\section{$\mathfrak{H}$-Shear-Free Null Geodesic Congruences}

\subsection{The Good Cut Equation on $\mathfrak{H}$}

In the study of $\mathfrak{I}^{+}$, several fascinating geometric structures
(such as $\mathcal{H}$-space and an asymptotic CR geometry) as well as
asymptotic physical identifications were discovered by considering Null
Geodesic Congruences (NGCs) whose shear was asymptotically vanishing \cite%
{UCF,PhysicalContent,Review}. \ The analogue of such a condition on a
non-expanding horizon is to look for those NGCs living exterior to $%
\mathfrak{H}$ whose shear vanishes at the intersection with the horizon. \
We refer to such NGCs as "$\mathfrak{H}$-shear-free," and the remainder of
this section will be devoted towards their study.

We have at this point a fixed null tetrad $\{l,n,m,\bar{m}\}$ on the NEH, $%
\mathfrak{H}$, where the vector $n$ is the only null vector pointing "off"
the horizon (i.e., $n$ is the only vector with a component in the $s$%
-direction). \ We now search for other null tetrad systems $\{l^{\ast
},n^{\ast },m^{\ast },\bar{m}^{\ast }\}$ (at $\mathfrak{H}$) that leave $%
l=l^{\ast }$ but force\ $n^{\ast }$ to be shear-free. Now, the shear of the $%
n$ congruence at $\mathfrak{H}$ is given by $-\bar{\lambda}$ (with $\lambda
=\lambda _{0}(\zeta ,\bar{\zeta})$); hence, for an NGC to be $\mathfrak{H}$%
-shear-free, we must be able to transform to a tetrad frame where $\lambda
^{\ast }\equiv 0$.

Using null rotations of the form (\ref{nullrotl}):%
\begin{eqnarray}
l &\rightarrow &l^{\ast }=l,  \label{nullrot*} \\
m &\rightarrow &m^{\ast }=m-Ll,  \notag \\
n &\rightarrow &n^{\ast }=n-\overline{L}m-L\bar{m}+L\overline{L}l,  \notag
\end{eqnarray}%
$\lambda $ transforms as \cite{Prior}:%
\begin{equation}
\lambda \rightarrow \lambda ^{\ast }=\lambda -\bar{L}\pi +2\bar{L}%
^{2}\varepsilon +\bar{L}^{2}\rho -\bar{L}^{3}\kappa +\bar{\eth }\bar{L}+\bar{%
L}D\bar{L}.  \label{ltrans}
\end{equation}%
Simplifying by using our tetrad conditions (\ref{tc1})-(\ref{tc3}), the
shear-free requirement ($\lambda ^{\ast }=0$) yields the $\mathfrak{H}$%
-shear-free condition:%
\begin{equation}
\eth L+L\dot{L}=-\bar{\lambda}_{0}(\zeta ,\bar{\zeta}),  \label{shearfree}
\end{equation}%
where $\dot{L}=\partial _{u}L$. \ 

\textbf{Remark: }Equation (\ref{shearfree}) is the same as the
asymptotically shear-free condition \cite{Aronson}:%
\begin{equation*}
\eth L+L\dot{L}=\sigma ^{0}(u,\zeta ,\bar{\zeta}),
\end{equation*}%
where $\sigma ^{0}$ is the asymptotic (Bondi) shear at $\mathfrak{I}^{+}$. \ 

To continue we assume that we are dealing with analytic functions or
functions that can be well approximated by analytic functions. We complexify 
$\mathfrak{H}$ by allowing $u$ to take on complex values close to the real
and $\bar{\zeta}$ to be independent of, but close to, the complex conjugate
of $\zeta .$

To solve Eq.(\ref{shearfree}) we transform it - via implicit differentiation
- into a simple 2$^{nd}$ order equation. A complex potential-like function $%
T(u,\zeta ,\bar{\zeta})$ is introduced, with level-surface values labeled by
the complex parameter $\tau $ (e.g., \cite{UCF}), 
\begin{equation}
\tau =T(u,\zeta ,\bar{\zeta}).  \label{T}
\end{equation}%
$T(u,\zeta ,\bar{\zeta})$ is defined from the $L(u,\zeta ,\bar{\zeta})$ by
solutions to the equation 
\begin{equation}
\eth _{(u)}T+L\dot{T}=0.  \label{CR1*}
\end{equation}%
The subscript $(u)$ in $\eth _{(u)}$ denotes the application of the $\eth $
operator while $u$ is held constant. Later we use $\eth _{(\tau )},$ with
the analogous meaning. \ Note that \ref{CR1*} is a CR equation and that both 
$T$ and $\bar{\zeta}$ are CR functions that determine a CR structure from $L$
on $\mathfrak{H.}$ This is in fact the same CR equation one obtains on $%
\mathfrak{I}^{+}$ \cite{ScrCr}. \ (See Appendix for details.)

We assume that the relationship $\tau =$ $T$ can be inverted to give%
\begin{equation}
u=G(\tau ,\zeta ,\bar{\zeta}),  \label{G}
\end{equation}%
which for constant $\tau $ gives a "slicing" of complex $\mathfrak{H}$ and
is referred to as a "good cut function." Using this inversion to change the
independent variable from $u$ to $\tau $ in equations (\ref{shearfree}) and (%
\ref{CR1*}), where in several steps implicit differentiation is used (c.f., 
\cite{UCF,Review}), we find, from (\ref{CR1*}), the relationship between the
good cut-function $G$ and the transformation function $L$,%
\begin{equation}
L=\eth _{(\tau )}G,  \label{ethG}
\end{equation}%
and from\ (\ref{shearfree}) the final relation, the "good cut equation"
itself \cite{PhysicalContent,Review}:%
\begin{equation}
\eth _{(\tau )}^{2}G(\tau ,\zeta ,\bar{\zeta})=-\bar{\lambda}_{0}(\zeta ,%
\bar{\zeta}).  \label{GCEq}
\end{equation}%
\qquad 

Before solving Eq.(\ref{GCEq}) several remarks are in order.

(1). \ The solution to good cut equation, Eq.(\ref{GCEq}), namely $u=G(\tau
,\zeta ,\bar{\zeta}),$ then gives us parametrically the solution to (\ref%
{shearfree}) by%
\begin{eqnarray}
L(u,\zeta ,\bar{\zeta}) &=&\eth _{(\tau )}G(\tau ,\zeta ,\bar{\zeta}),
\label{parametric} \\
u &=&G(\tau ,\zeta ,\bar{\zeta}).  \notag
\end{eqnarray}

(2). There is a very pretty geometric meaning of the null rotation function $%
L(u,\zeta ,\bar{\zeta}).$ The sphere of null generators of the past
light-cone of each point of $\mathfrak{H}$ can be labeled by a stereographic
angle with the infinity generator, ($L=\infty $) lying on $\mathfrak{H.}$
The function $L(u,\zeta ,\bar{\zeta})$ is the stereographic angle field
giving the null directions of the $\mathfrak{H}$-shear-free vector field $%
n^{\ast }$ at its intersection with $\mathfrak{H}.$

\subsection{Solving the Good Cut Equation on $\mathfrak{H}$}

Following closely the analogy with the study of asymptotically shear-free
NGCs, we show that the solutions to (\ref{GCEq}); depend only on four
complex parameters, i.e. the solution space is a four-complex dimensional
manifold. \ In the case where $\eth =\eth _{0}$ (i.e., where the $u=const.$
cross-sections are 2-spheres), this follows easily from the properties of
the $\eth _{0}$-operator on 2-spheres and its operation on spin-weight $s$
tensorial spherical harmonics. \ For a general non-expanding horizon
however, the situation is slightly more complicated, since $\eth $ acts on a
2-manifold which is only \textit{conformal} to a 2-sphere.

We first show that in the homogeneous case (i.e., where $\lambda _{0}=0$),
the good cut equation on $\mathfrak{H}$ has a complex four-dimensional
solution space. \ Using this fact, we subsequently prove that this remains
true for the general case of (\ref{GCEq}). We conclude the sub-section with
the observation that regular $\mathfrak{H}$-shear-free NGCs are thus
generated by complex world-lines in this complex four-manifold.

To begin, consider the homogeneous good cutequation on $\mathfrak{H}$:%
\begin{equation}
\eth ^{2}G_{0}=0.  \label{HGCE}
\end{equation}%
Using (\ref{edth}), (\ref{ethG}), and (\ref{conformal}), as well as
recalling that $G_{0}$ (the general homogeneous solution) is a spin-weight
zero function, this can be re-written as:%
\begin{eqnarray}
\eth ^{2}G_{0} &=&\eth \left( P\frac{\partial }{\partial \zeta }G_{0}\right)
=\eth \left( V\eth _{0}G_{0}\right) =0  \label{h2} \\
&=&\frac{\partial }{\partial \zeta }P\left( V\eth _{0}G_{0}\right) =\frac{%
\partial }{\partial \zeta }P_{0}V\left( V\eth _{0}G_{0}\right) =0  \notag \\
&=&\eth _{0}G_{0}\cdot \eth _{0}V^{2}+V^{2}\eth _{0}^{2}G_{0}=0.  \notag
\end{eqnarray}%
Writing the spin-weight one function $\eth _{0}G_{0}$ as%
\begin{equation}
F\equiv \eth _{0}G_{0},  \label{h3}
\end{equation}%
after some algebraic manipulation, we obtain:%
\begin{equation}
F^{-1}\eth _{0}F+V^{-2}\eth _{0}V^{2}=\eth _{0}[\log (FV^{2})]=0.  \label{h4}
\end{equation}%
In turn this implies 
\begin{equation}
\eth _{0}(FV^{2})=0.  \label{h5}
\end{equation}%
Since we have now reduced (\ref{HGCE}) to a relation involving only the
2-sphere operator, we can easily solve (\ref{h5}), using the properties of $%
\eth _{0}$ and the tensorial spin weight-$s$ spherical harmonics, as:%
\begin{equation}
FV^{2}=z^{i}Y_{1i}^{1}(\zeta ,\bar{\zeta}),  \label{h6}
\end{equation}%
where the $z^{i}$ are three arbitrary complex parameters. \ Eq.(\ref{h3}) is
thus:%
\begin{equation}
\eth _{0}G_{0}=V^{-2}z^{i}Y_{1i}^{1},  \label{h7}
\end{equation}%
with the general solution%
\begin{equation}
G_{0}(z^{a},\zeta ,\bar{\zeta})=z^{0}+\doint\limits_{S^{2}}K_{1}(\zeta ,\bar{%
\zeta};\eta ,\bar{\eta})V^{-2}z^{i}Y_{1i}^{1}\ dS_{1}.  \label{h8}
\end{equation}%
$K_{1}$ is a known Green's function for the $\eth _{0}$-operator and the
measure on $S^{2}$ is \cite{Greens}:%
\begin{equation}
dS_{1}=-2i\frac{d\eta \wedge d\bar{\eta}}{(1+\eta \bar{\eta})^{2}}.
\label{h9}
\end{equation}

We thus see that the homogeneous good cut equation on $\mathfrak{H}$ does
indeed depend on four complex parameters, denoted $z^{a}\in \mathbb{C}$. \
The question remains: does this generalize to the fully inhomogeneous case?
\ The answer is in the affirmative. \ Let us assume that the general
solution to the (inhomogeneous) good cut equation can be written as 
\begin{equation}
G=G_{0}\mathfrak{g},  \label{i1}
\end{equation}%
for some undetermined function $\mathfrak{g}$. \ Then we have, using Eq.(\ref%
{HGCE}),%
\begin{equation}
\eth ^{2}(G_{0}\mathfrak{g})=G_{0}\eth ^{2}\mathfrak{g}+2\eth G_{0}\eth 
\mathfrak{g=}-\bar{\lambda}_{0},  \label{i2}
\end{equation}%
which, using%
\begin{equation}
\mathfrak{f}\equiv \eth \mathfrak{g},  \label{i3}
\end{equation}%
reduces to:%
\begin{equation}
\eth (G_{0}^{2}\mathfrak{f})=-G_{0}\bar{\lambda}_{0}.  \label{i4}
\end{equation}%
Again recalling the relationship between the $\eth $ and $\eth _{0}$%
-operators, this is rewritten as%
\begin{equation}
\eth _{0}\left( VG_{0}^{2}\mathfrak{f}\right) =-G_{0}\bar{\lambda}_{0},
\label{i5}
\end{equation}%
implying that%
\begin{equation}
\mathfrak{f}=\eth \mathfrak{g}=-V^{-1}G_{0}^{-2}\doint\limits_{S^{2}}K_{2}(%
\zeta ,\bar{\zeta};\eta ,\bar{\eta})G_{0}\bar{\lambda}_{0}dS_{2}.  \label{i6}
\end{equation}%
Noting that $\eth \mathfrak{g}=V\eth _{0}\mathfrak{g}$, we obtain%
\begin{equation}
\mathfrak{g}=-\doint\limits_{S^{2}}\left( K_{1}(\zeta ,\bar{\zeta};\eta _{2},%
\bar{\eta}_{2})V^{-2}(\eta _{2},\bar{\eta}_{2})G_{0}^{-2}(\eta _{2},\bar{\eta%
}_{2})\doint\limits_{S^{2}}K_{2}(\eta _{2},\bar{\eta}_{2};\eta _{1},\bar{\eta%
}_{1})G_{0}(\eta _{1},\bar{\eta}_{1})\bar{\lambda}_{0}(\eta _{1},\bar{\eta}%
_{1})dS_{1}\right) dS_{2}.  \label{i7}
\end{equation}

Combining this with equation (\ref{i1}), we see that the general solution $%
G(\tau ,\zeta ,\bar{\zeta})$ depends only on the shear data $\lambda
_{0}(\zeta ,\bar{\zeta})$ and the four complex parameters of the homogeneous
solution, $z^{a}\in \mathbb{C}$. \ It indeed follows that solutions to the
good cut equation on $\mathfrak{H}$ live in a complex four dimensional
manifold in the same fashion as solutions to the asymptotic good cut
equation!

Finally to find the solutions with $\tau $ dependence all that must be done
is to replace the $z^{a}$ by an arbitrary analytic function of $\tau ,$
i.e., by $z^{a}(\tau ).$ Arbitrary analytic curves in the associated complex
four dimensional space generate solutions to the good cut equation which in
turn generate, via the parametric relations, Eq.(\ref{parametric}), $%
\mathfrak{H}$-shear-free null geodesic congruences in the neighborhood of $%
\mathfrak{H.}$ As mentioned earlier, it follows that every choice of such
world-line induces a CR structure on the horizon (see Appendix).

It should be noted that if the ambient space-time is algebraically special
then it - the space-time itself - possesses\ a unique shear-free congruence
which is also $\mathfrak{H}$-shear-free thereby endowing $\mathfrak{H}$ with
a unique CR structure. \ For more general space-times, one can ask: is there
a way to canonically single out a particular choice of world-line, thereby
choosing a unique $\mathfrak{H}$-shear-free CR structure for any vacuum NEH?

\subsection{Unique Choice of World-line}

In this final sub-section, we describe a method for singling out a unique
world-line $z^{a}(\tau )$ which in turn generates a unique solution to the
good cut equation (\ref{GCEq}) on $\mathfrak{H}$ and a related unique CR
structure. \ Again, we take our motivation from the extant physical
identification theory on $\mathfrak{I}^{+}$, which uses similar concerns to
single out a unique complex world-line in $\mathcal{H}$-space (\cite%
{UCF,PhysicalContent,Review}). \ In the asymptotic theory, a unique
world-line is singled out by transforming to a tetrad frame where the
complex gravitational dipole moment (proportional to the $l=1$ spherical
harmonic contribution to $\psi _{1}^{0}$) vanishes. \ In order to do a
similar transformation on a vacuum NEH, we must first identify what will
serve as our complex gravitational dipole at $\mathfrak{H}$.

The best choice appears to be the $l=1$ contribution from $\bar{\psi}_{3}$,
which has the proper spin-weight ($s=1$) and transformation behavior to
serve as a base for the description of a dipole. \ Performing a spherical
harmonic expansion%
\begin{equation*}
\bar{\psi}_{3}=\bar{\psi}_{3}^{i}Y_{1i}^{1}+\bar{\psi}_{3}^{ij}Y_{2ij}^{1}+%
\cdots ,
\end{equation*}%
we write%
\begin{equation}
\bar{\psi}_{3}^{i}=-\frac{6\sqrt{2}G}{c^{2}}D_{\mathbb{C}}^{i}=-\frac{6\sqrt{%
2}G}{c^{2}}(D_{(mass)}^{i}+ic^{-1}J^{i}),  \label{Dipole}
\end{equation}%
where $D_{\mathbb{C}}^{i}$ is identified as (some form) of a complex
`gravitational' dipole (c.f., \cite{Review}). \ The process of singling out
a unique world-line then becomes the task of transforming to a tetrad frame
where $D_{\mathbb{C}}^{i}$, or equivalently the new $\bar{\psi}_{3}^{i\ast }$%
, vanishes. \ To do this, we perform a null rotation about the null vector $%
l $ using the function $L=\eth G$ from (\ref{parametric}). \ Under such a
null rotation, the Weyl tensor component $\bar{\psi}_{3}$ transforms as \cite%
{Prior}:%
\begin{equation*}
\bar{\psi}_{3}\rightarrow \bar{\psi}_{3}^{\ast }=\bar{\psi}_{3}-3L\bar{\psi}%
_{2}.
\end{equation*}%
The "center of mass" condition, $\bar{\psi}_{3}^{i\ast }=0,$ thus leads to:%
\begin{equation}
0=\bar{\psi}_{3}^{i}-3(L\bar{\psi}_{2})|^{i},  \label{CM}
\end{equation}%
where \TEXTsymbol{\vert}$^{i}$ means, "extract the $l=1$ part of the
product". \ Now, both $\psi _{3}$ and $\psi _{2}$ are quantities given in
terms of the free data on $\mathfrak{H}$, while the function $L$ carries the
information about the good cut function via the complex world-line $%
z^{a}(\tau )$. \ Hence, (\ref{CM}) is an algebraic equation for the choice
of complex world-line in terms of the free data on the horizon. \ 

This means that we not only single out a particular complex world-line as
seen from $\mathfrak{H}$, but also obtain a unique solution to the good cut
equation, which induces a unique CR structure on the horizon! \ 

\section{Discussion and Conclusion}

In this paper we have set out to do several things. \ First, we returned to
the old topic of non-expanding horizons, and gave a complete description of
their vacuum geometry using the spin-coefficient formalism. \ In particular,
we exploited the gauge freedom to construct a unique null tetrad and,
essentially, a unique choice of coordinates on the horizon. In addition we
reduced the amount of free data on the horizon to five complex and three
real quantities (see sub-section 2.2). \ Besides generalizing the earlier
work of Pajerski and extending the more recent results of Ashtekar and
others with the spin-coefficient formalism, this also provided us with a
platform for investigating the geometric structure induced on NEHs by
considering $\mathfrak{H}$-shear-free null geodesic congruences.

In Section 3, we demonstrated that looking for such $\mathfrak{H}$%
-shear-free congruences results in a good cut equation (\ref{GCEq}) that
closely resembles the well-studied asymptotic good cut equation on $%
\mathfrak{I}^{+}$. \ Additionally, it was shown that solutions to this
equation lie in a complex four-dimensional space, and that arbitrary choices
of world-lines in this space generate analytic $\mathfrak{H}$-shear-free
null geodesic congruences at the horizon. \ We also showed that by making an
identification of a complex dipole term, it is possible to single out a
unique such world-line that corresponds - in some sense - to the complex
center of mass. \ This particular world-line would then induce a unique
solution to the good cut equation and thus allow a unique CR structure on
the horizon associated with the $\mathfrak{H}$-shear-free null geodesic
congruence. \ This unique good cut function is the direct analogue of the
Universal Cut Function on the $\mathfrak{I}^{+}$ of asymptotically flat
space-times, which is derived in a similar fashion \cite{UCF,Review}.

Through much of this paper, we have motivated many of our calculations using
the analogy between a NEH $\mathfrak{H}$ and future null infinity $\mathfrak{%
I}^{+}$. \ The next step one would like to take in this analogy would be to
construct a means of physical identification on non-expanding horizons using
the same tools as in asymptopia. \ It is at this point where the analogy may
break down. \ It is not at all clear what is the relationship - if any -
between the four-dimensional solution space of the good cutequation, when $P$
does not represent a sphere metric, and the $\mathcal{H}$-space arising from
the $P_{0}.$ An immediate question would be: does this new four- dimensional
space define a complex metric analogous to that of $\mathcal{H}$-space?
This, as well as the issue of giving a physical identification of the
world-line, are open questions being investigated.

Despite these difficulties, it should be noted that for the special case
where $u=const.$ cross sections of $\mathfrak{H}$ have a metric factor $%
P=P_{0}$ (i.e., spheres), the ambiguities just mentioned largely disappear,
and it might be possible to proceed with a physical identification theory. \ 

A final question that is raised by our work here regards the manifold in
which the world-lines generating $\mathfrak{H}$-shear-free null geodesic
congruences live. \ For $\mathfrak{I}^{+}$, a surprising metric construction 
\cite{HSM} allowed such world-lines to be interpreted as lying in a complex
Minkowski space or $\mathcal{H}$-space. \ We hope that the solution manifold
that appears for horizons is isomorphic (or related in some sense) to $%
\mathcal{H}$-space, or at least a complex deformation away from it, but this
is currently an open research question.

\section{Appendix}

\subsection{CR Structures on $\mathfrak{H}$}

A CR structure on a real three manifold $\mathfrak{H}$ , with local
coordinates $x^{a}$, is given intrinsically by equivalence classes of
1-forms: one real, one complex and its complex conjugate \cite{CR1}. \ If we
denote the real 1-form by $\mathfrak{L}$ and the complex one-form by $%
\mathfrak{M}$, then these are defined up to the transformations:%
\begin{eqnarray}
\mathfrak{L} &\rightarrow &a(x^{a})\mathfrak{L},  \label{Trans1} \\
\mathfrak{M} &\rightarrow &f(x^{a})\mathfrak{M}+g(x^{a})\mathfrak{L}.  \notag
\end{eqnarray}%
The $(a,f,g)$ are functions on $\mathfrak{H}$: $a$ is non-vanishing and
real, $f$ and $g$ are complex function with $\ f$ non-vanishing. \ Further
it is required that there be a three-fold linear independence relation
between these 1-forms \cite{CR1}:%
\begin{equation}
\mathfrak{L}\wedge \mathfrak{M}\wedge \mathfrak{\bar{M}}\neq 0.  \label{Lin}
\end{equation}

Any three-manifold with a CR structure is referred to as a three-dimensional
CR manifold. \ There are special classes (referred to as embeddible) of 3-D
CR manifolds that can be directly embedded into $\mathbb{C}^{2}$. \ For the
NEH $\mathfrak{H}$, we have the differential equation - the so-called CR
equation,%
\begin{equation}
\eth K+L\dot{K}\equiv P\frac{\partial K}{\partial \zeta }+L\frac{\partial K}{%
\partial u}=0,  \label{CR1}
\end{equation}%
where the two linearly independent solutions $K_{1,2}:\mathfrak{H}%
\rightarrow \mathbb{C}$ give the embedding of the NEH into $\mathbb{C}^{2}$.
\ The first of these solutions is rather obvious:%
\begin{equation}
K_{1}(u,\zeta ,\bar{\zeta})=\bar{\zeta},  \label{CRF1}
\end{equation}%
and the second is given by the complex potential function (see Eq.(\ref{CR1*}%
))%
\begin{equation}
K_{2}(u,\zeta ,\bar{\zeta})=T(u,\zeta ,\bar{\zeta})=\tau .  \label{CRF2}
\end{equation}

To prove that (\ref{CRF1}) and (\ref{CRF2}) indeed describe a CR structure
on $\mathfrak{H}$, we must show that we can derive the class of 1-forms (\ref%
{Trans1}). \ 

This is done in the following manner: let $(\tilde{\tau},\tilde{\zeta})$ be
coordinates on $\mathbb{C}^{2}$; when these coordinates are restricted to $%
\mathfrak{H}\subset \mathbb{C}^{2}$, we get:%
\begin{equation*}
\tilde{\tau}|_{\mathfrak{H}}=\tau ,\ \ \tilde{\zeta}|_{\mathfrak{H}}=\bar{%
\zeta}=x-iy.
\end{equation*}%
Now, taking the exterior derivatives of these quantities (restricted to $%
\mathfrak{H}$) gives us two (complex) 1-forms on the horizon:%
\begin{eqnarray}
\mathfrak{K}_{1} &=&d\bar{\zeta},  \label{CR3} \\
\mathfrak{K}_{2} &=&d\tau =\frac{\partial T}{\partial x^{a}}dx^{a}
\label{CR3*} \\
&=&\dot{T}du+\frac{\partial T}{\partial \zeta }d\zeta +\frac{\partial T}{%
\partial \bar{\zeta}}d\bar{\zeta},  \notag
\end{eqnarray}%
where $\mathfrak{K}_{1}$ and $\mathfrak{K}_{2}$ are defined up to
bi-holomorphic transformations on the coordinates $\tilde{\tau}$ and $\tilde{%
\zeta}$ of $\mathbb{C}^{2}$ \cite{CR2}. \ That is, we retain the freedom:%
\begin{eqnarray}
\mathfrak{K}_{1} &\rightarrow &f(x^{a})\mathfrak{K}_{1}+g(x^{a})\mathfrak{K}%
_{2},  \label{CR4} \\
\mathfrak{K}_{2} &\rightarrow &h(x^{a})\mathfrak{K}_{1}+j(x^{a})\mathfrak{K}%
_{2}.  \notag
\end{eqnarray}%
We can use (\ref{CR4}) to make $\mathfrak{K}_{2}$ a real 1-form by dividing (%
\ref{CR3*}) by $\dot{T}$ and then using (\ref{CR1*}) to obtain:%
\begin{equation*}
\mathfrak{K}_{2}\rightarrow \mathfrak{L}=du-\frac{L}{P}d\zeta -\frac{\bar{L}%
}{P}d\bar{\zeta}.
\end{equation*}%
\qquad 

This reduces the remaining freedom in the forms to (\ref{Trans1}) and gives
the required pair of 1-forms (one real, one complex):%
\begin{eqnarray*}
\mathfrak{L} &=&du-\frac{L}{P}d\zeta -\frac{\bar{L}}{P}d\bar{\zeta}, \\
\mathfrak{M} &=&d\bar{\zeta},
\end{eqnarray*}%
which define a CR structure on $\mathfrak{H}$, as required.

Note that for each $L$ (i.e., for each choice of complex world-line $%
z^{a}(\tau )$), we obtain a different CR structure. \ In particular, for the
world-line corresponding to the complex center of mass via equation (\ref{CM}%
), there is a unique CR structure on the horizon generated by the "complex
center of mass."

\end{document}